\documentclass[twocolumn,showpacs,preprintnumbers,superscriptaddress,amsmath,amssymb,pre]{revtex4}
\usepackage{amsmath}

\usepackage{graphicx}% Include figure files
\usepackage{dcolumn}% Align table columns on decimal point
\usepackage{bm}% bold math

\begin{document}

%\preprint{Submitted to {\em{Physical Review letter}}}

\title{Direct evidence for inversion formula in multifractal financial volatility measure}

\author{Zhi-Qiang Jiang}
 \affiliation{School of Business, East China University of Science and Technology, Shanghai 200237, China}
 \affiliation{School of Science, East China University of Science and Technology, Shanghai 200237, China}

\author{Wei-Xing Zhou}
 \email{wxzhou@ecust.edu.cn}
 \affiliation{School of Business, East China University of Science and Technology, Shanghai 200237, China}
 \affiliation{School of Science, East China University of Science and Technology, Shanghai 200237, China}
 \affiliation{Research Center for Econophysics, East China University of Science and Technology, Shanghai 200237, China}
 \affiliation{Research Center of Systems Engineering, East China University of Science and Technology, Shanghai 200237, China}

\date{\today}
%%%%%%%%%%%%%%%%%%%%%%%%%%%%%%

\begin{abstract}
The inversion formula for conservative multifractal measures was
unveiled mathematically a decade ago, which is however not well
tested in real complex systems. In this Letter, we propose to verify
the inversion formula using high-frequency turbulent financial data.
We construct conservative volatility measure based on minutely S\&P
500 index from 1982 to 1999 and its inverse measure of exit time.
Both the direct and inverse measures exhibit nice multifractal
nature, whose scaling ranges are not irrelevant. Empirical
investigation shows that the inversion formula holds in financial
markets.
\end{abstract}


\maketitle

In recent years, the concept of inverse statistics has attracted
much attention in turbulence
\cite{Jensen-1999-PRL,Biferale-Cencini-Vergni-Vulpiani-1999-PRE} as
well as in financial markets
\cite{Simonsen-Jensen-Johansen-2002-EPJB} based on time series
analysis. The direct structure function concerns with the
statistical moments of a physical quantity $g$ measured over a
distance $s$ such that $S_q(s) = \langle g_\| (s)^q \rangle$. The
multifractal nature of direct structure functions has been well
documented in turbulence
\cite{McCauley-1990-PR,Frisch-1996,Anselmet-Gagne-Hopfinger-Antonia-1984-JFM},
as well as in finance
\cite{Vandewalle-Ausloos-1998-EPJB,Ivanova-Ausloos-1999-EPJB,Calvet-Fisher-2002-RES},
which is characterized by $S_q(s) \sim s^{\zeta(q)}$ with a
nonlinear scaling function $\zeta(q)$. In contract, the inverse
structure function is related to the exit distance, where the
physical quantity fluctuation exceeds a prescribed value, such that
$T_p(g) = \langle s^p(g) \rangle$. One can intuitively expect that
there is a power law scaling stating that $T_p(g) \sim g^{\phi(p)}$,
where $\phi(p)$ is also a nonlinear function. Furthermore, if $s\sim
g^{1/h}$, Schmitt has shown that there is an inversion formula
between the two types of scaling exponents such that $\zeta(q)=-p$
and $\phi(p)=-q$ \cite{Schmitt-2005-PLA}. A similar intuitive
derivation for the inversion formula is given for Laplacian random
walks \cite{Hastings-2002-PRL}.

The power-law scaling in inverse structure function was observed in
the signals of two dimensional turbulence
\cite{Biferale-Cencini-Lanotte-Vergni-Vulpiani-2001-PRL,Biferale-Cencini-Lanotte-Vergni-2003-PF},
in the synthetic velocity data of the GOY shell model
\cite{Jensen-1999-PRL,Roux-Jensen-2004-PRE}, and in the temperature
and longitudinal and transverse velocity data in grid-generated
turbulence \cite{Beaulac-Mydlarski-2004-PF}. However, this scaling
behavior was not observed in other three dimensional turbulent flows
from different experiments
\cite{Biferale-Cencini-Vergni-Vulpiani-1999-PRE,Pearson-vandeWater-2005-PRE,Zhou-Sornette-Yuan-2006-PD}.
The inversion formula for direct and inverse structure functions is
verified for synthetic turbulence data of shell models
\cite{Roux-Jensen-2004-PRE} but not for wind-tunnel turbulence data,
which cover a range of Reynolds numbers ${\rm{Re}} = 400-1000$
\cite{Pearson-vandeWater-2005-PRE}.

It is argued that \cite{Xu-Zhou-Liu-Gong-Wang-Yu-2006-PRE}, the
absence of inversion formula between the scaling exponents of direct
and inverse structure functions is due to the facts that the
velocity fluctuation is not a conservative quantity while a strict
proof of the inversion formula was given for conservational
multifractal measures
\cite{Mandelbrot-Riedi-1997-AAM,Riedi-Mandelbrot-1997-AAM}. It is
noteworthy pointing out that, the inversion formula given by Roux
and Jensen is obtained based on a special case of conservative
measures, although they verified the inversion formula in the direct
and inverse structure functions of shell models
\cite{Roux-Jensen-2004-PRE}. It is thus natural that Xu {\em{et
al.}} proposed to test the inversion formula in the energy
dissipation rate (a kind of conservative measure) rather than in the
structure functions and they did found sound evidence in favor of
the proof \cite{Xu-Zhou-Liu-Gong-Wang-Yu-2006-PRE}.

The inversion formula was theoretically established for both
discontinuous and continuous multifractal measures by Riedi and
Mandelbrot
\cite{Mandelbrot-Riedi-1997-AAM,Riedi-Mandelbrot-1997-AAM}. Let
$\mu$ be a probability measure on $[0,1]$ whose integral function
$M(a)=\mu([0,a])$ is right-continuous and nondecreasing. Since the
measure is self-similar, we have $\mu =
\sum_{i=0}^{n}m_i\mu(w_i^{-1}(\cdot))$, where $w_i$'s are the
similarity maps with scale contraction ratios $r_i \in (0,1)$ and
$\sum_{i=1}^n m_i = 1$ with $m_i > 0$. The multifractal spectrum
$f(\alpha)$ of measure $\mu$ can be obtained via the Legendre
transform of $\tau$, which is defined by
\begin{equation}
\sum_{i=1}^n m_i^qr_i^{-\tau} = 1~.
 \label{Eq:dtau}
\end{equation}
The inverse measure of $\mu^*$ can be defined as follows,
\begin{equation}
\mu^* = M^*(b)=\left\{
\begin{array}{lll}
\inf \{a:M(a)>b\} && {\rm{if}}~b < 1 \\
1 && {\rm{if}}~ b=1
\end{array}
\right., \label{Eq:inversemeasure}
\end{equation}
where $M^*(b)$ is the inverse function of $M(a)$. Since $\mu$ is
self-similar, its inverse measure $\mu^*$ is also self-similar with
ratios $r_i^* = m_i$ and probabilities $m_i^* = r_i$, whose
multifractal spectrum $f^*(\alpha^*)$ is the Legendre transform of
$\theta$, which is defined implicitly by
\begin{equation}
\sum_{i=1}^n (m^*_i)^p (r_i^*)^{-\theta} = 1~.
 \label{Eq:dtheta}
\end{equation}
The inversion formula follows immediately that
\begin{equation} \label{Eq:IF}
\left\{
\begin{array}{lll}
  \tau(q) &=& -p \\
  \theta(p) &=& -q
\end{array} \right..
\end{equation}
Equivalently, we have
\begin{equation}
\tau(q) = -\theta^{-1}(-q)
 \label{Eq:IF1}
\end{equation}
or
\begin{equation}
\theta(p) = -\tau^{-1}(-p)~.
 \label{Eq:IF2}
\end{equation}

These two equivalent relations are testable. Following this line,
the inversion formula was verified in
\cite{Xu-Zhou-Liu-Gong-Wang-Yu-2006-PRE} with high-Reynolds
turbulence data collected at the S1 ONERA wind tunnel
\cite{Anselmet-Gagne-Hopfinger-Antonia-1984-JFM}, which is however
the only evidence. Due to the well documented analogues between
turbulent flows and financial markets \cite{Mantegna-Stanley-2000},
in this letter, we propose to test the inversion formula in
financial markets using high-frequency historical data of the S\&P
500 index. Our data consist of 18-year minutely prices spanning from
1 January, 1982 to 31 December, 1999 with a total of 1.7 million
data points. The minutely return $r(t)$ is calculated as follows,
\begin{equation}
r(t) = \ln [I(t)/I(t-1)]~,
 \label{Eq:return}
\end{equation}
where $\{I(t):t=1,\cdots,T\}$ is the time series of minutely S\&P
500 index.

We first construct the direct volatility measure and investigate its
multifractal nature. The absolute return is utilized as a proxy for
volatility such that $v(t) = |r(t)|$. According to the partition
function method for multifractal analysis, the series is firstly
covered by $N$ boxes with identical size $s=T/N$. The sizes of the
boxes are chosen such that the number of boxes of each size is an
integer to cover the whole time series. On each box, we construct
the direct measure as
\begin{equation}
\mu_n(s) = \frac{1}{V}\sum_{t=(n-1)s+1}^{ns} v(t),
 \label{Eq:dmeasure}
\end{equation}
where $V=\sum_{t=1}^{T}v(t)$ and $n=1,\cdots,N$. By construction,
this volatility measure $\mu$ is conservative. The presence of
multifractality in $\mu$ has been confirmed based on the multiplier
method utilizing the same data set of the minutely S\&P 500 index
\cite{Jiang-Zhou-2007-PA}. Alternatively, the volatility measure
$\mu$ of Chinese stocks and indexes exhibits multifractal behavior
based on the partition function approach \cite{Jiang-Zhou-2008-XXX}.

For order $q$, the direct partition function $\chi_q(s)$ can be
estimated using
\begin{equation}
\chi_q(s) =  \sum_{n=1}^{N} [\mu_n(s)]^q~.
 \label{Eq:dpf}
\end{equation}
When $\mu \ll 1$ and $q \gg 1$, the estimation of the partition
function $\chi$ will be very difficult since the value is so small
that it is ``out of the memory''. To overcome this problem, we can
calculate the logarithm of the partition function $\ln \chi_q(s)$
rather than the partition function itself. A simple manipulation
results in the following formula
\begin{equation}
\ln \chi_q(s) = \ln \sum_{n=1}^{N} \left[
\frac{\mu_n(s)}{\mu_{\max}}\right]^q + q \ln \mu_{\max}~,
 \label{Eq:logchi}
\end{equation}
where $\mu_{\max}=\max\{\mu_n: n=1,\cdots,N\}$. This trick applies
for the calculation of inverse partition functions as well.

\begin{figure}[htb]
\centering
\includegraphics[width=6.5cm]{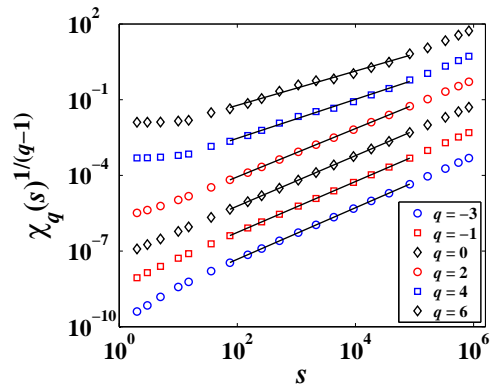}
\caption{(color online.) Dependence of $\chi_q(s)^{1/q-1}$ as a
function of box size $s$ for $q = -3$, $q = -1$, $q = 0$, $q =4$,
and $q = 6$. The curves have been translated vertically by a factor
of 0.001, 0.01, 0.1, 10, and 100 in turn for better visibility. The
solid lines are power-law fits in the scaling range.}
\label{Fig:PF:direct}
\end{figure}

Figure \ref{Fig:PF:direct} plots $[\chi_q(s)]^{1/q-1}$ as a function
of box size $s$ for different orders. Sound power laws are observed
for each partition function such that
\begin{equation}
\chi_q(s) \sim s^{\tau(q)}~,
 \label{Eq:chiq}
\end{equation}
in which the scaling range spans about three orders of magnitude.
The scaling exponent $\tau(q)$ can be estimated through a power-law
fit to the data in the scaling range. We will see that $\tau(q)$ is
a nonlinear function, confirming the presence of multifractality in
the direct volatility measure.

We now investigate the scaling behavior of inverse partition
function of exit times. For each threshold $\Delta{v}$, a sequence
of exit times $s_j(\Delta v)$ can be determined successively from
$j=1$ to $j=J$ by
\begin{equation}
 \sum_{k=1}^j s_k = \inf \left\{t : \int_0^t \nu(t)dt \geqslant j \Delta v \right\}~,
 \label{Eq:exittime}
\end{equation}
where $\nu(t)=v(t)$ for $t\in[t,t+1)$. The inverse measure is
defined as the normalized exit time
\begin{equation}
\mu^*_j(\Delta{v}) = s_j/ T~,
 \label{Eq:imeasure}
\end{equation}
and the inverse partition function can be determined as follows
\begin{equation}
\chi_p^*(\Delta{v}) =  \sum_{j=1}^J
\left[\mu^*_j(\Delta{v})\right]^p~,
 \label{Eq:iPF}
\end{equation}
where
\begin{equation}
  J=\left[\frac{1}{\Delta{v}}\int_0^T\nu(t)dt\right]~.
 \label{Eq:J}
\end{equation}

\begin{figure}[htb]
\centering
\includegraphics[width=6.5cm]{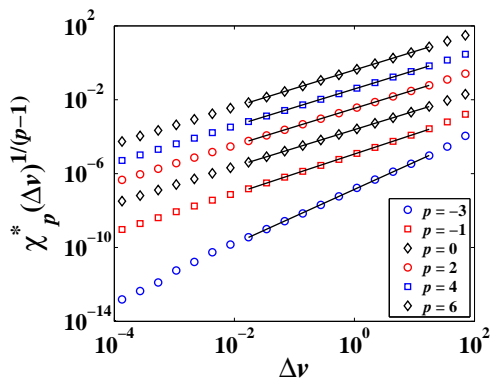}
\caption{(color online.) Dependence of $\chi^*_p(\Delta v)^{1/p-1}$
on the thresholds $\Delta v$. The same translated processes are also
performed on the data points of $p = -3$, $p = -1$, $p = 0$, $p =4$,
and $p = 6$. The solid lines are the best fits to the data.}
\label{Fig:PF:inverse}
\end{figure}

Figure~\ref{Fig:PF:inverse} shows the dependence of $\chi^*_p(\Delta
v)^{1/(p-1)}$ on the threshold values $\Delta{v}$ for different
values of $p$. Power-law scaling can be observed
\begin{equation}
\chi_p^*(s) \sim \Delta v^{\theta(p)}~.
 \label{Eq:chip2}
\end{equation}
where the scaling range covers about three orders of magnitude. The
straight lines are the best fits to the data in the scaling range,
whose slopes correspond to the exponents $\theta(p)/(p-1)$. We note
that the two scaling ranges $(s_1,s_2)$ and
$(\Delta{v}_1,\Delta{v}_2)$ for direct and inverse partition
functions are related by
\begin{equation}
 \Delta{v}=s\times{v}_{\rm{mean}}~,
\end{equation}
where $v_{\rm{mean}}=2.69\times10^{-4}$. Specifically, we find that
$\Delta{v}_1\approx{s_1}\times{v}_{\rm{mean}}$ and
$\Delta{v}_2\approx{s_2}\times{v}_{\rm{mean}}$. This puts forward
sound evidence upon the determination of the scaling laws.

A subtle issue concerning negative moments arises, which is related
to the probability density functions (PDFs) of volatility and exit
time respectively. The volatility of S\&P 500 index is log-normally
distributed in the center followed by a power-law tail for large
volatilities while the right tail seems truncated
\cite{Cizeau-Liu-Meyer-Peng-Stanley-1997-PA,Liu-Gopikrishnan-Cizeau-Meyer-Peng-Stanley-1999-PRE}.
In addition, due to the construction of the volatility measure,
$\mu(s)>0$ for large $s$ in the scaling range shown in
Fig.~\ref{Fig:PF:direct}. Therefore, negative moments can be
estimated numerically. Taking into account the statistical
significance of the estimation of partition functions
\cite{Lvov-Podivilov-Pomyalove-Procaccia-Vandembroucq-1998-PRE,Zhou-Sornette-Yuan-2006-PD},
we focus on $q\in[-4,8]$.

\begin{figure}[htb]
\centering
\includegraphics[width=6.5cm]{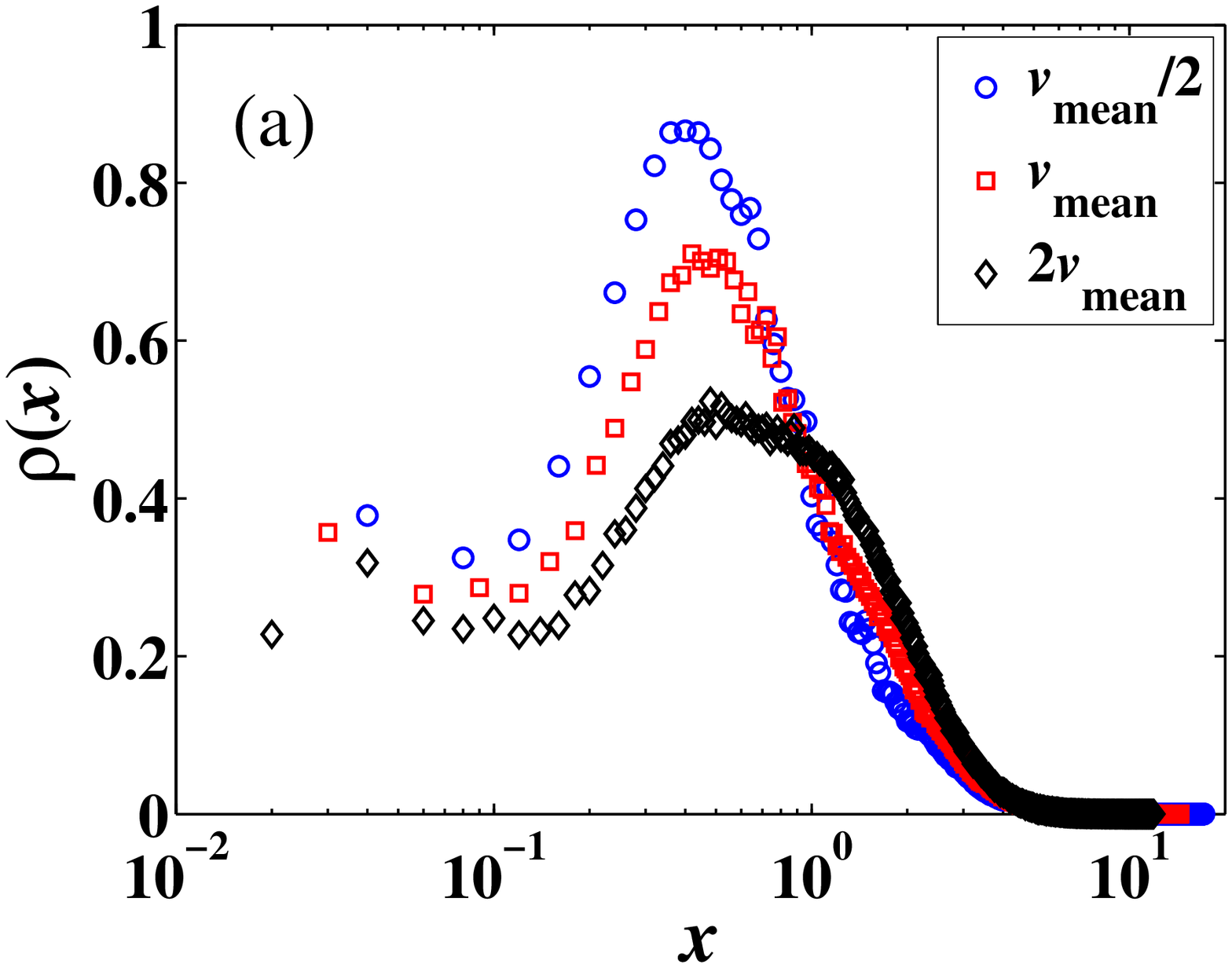}
\includegraphics[width=6.5cm]{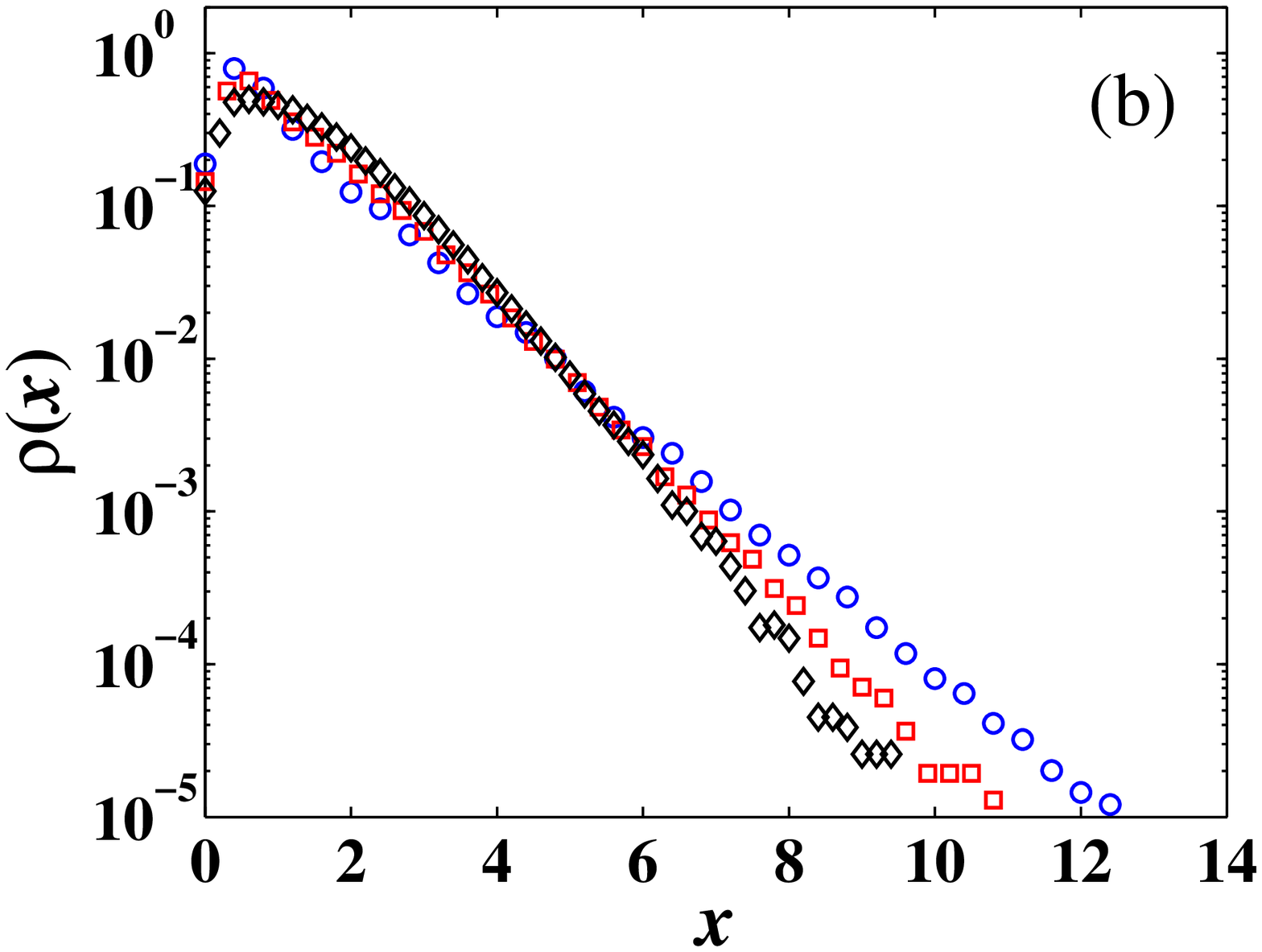}
\caption{(color online.) Empirical probability density function
$\rho(x)$ of the normalized exit time $x = s/\sigma$ for different
thresholds $\Delta v= v_{\rm{mean}}/2$ ($\circ$), $v_{\rm{mean}}$
($\Box$), and $2v_{\rm{mean}}$ ($\diamond$).} \label{Fig:PDF}
\end{figure}

The PDFs of exit time defined in this letter have not been
investigated before. Let us denote by $f(s)$ the PDF of exit times
for a fixed threshold $\Delta{v}$. For comparison, we normalize the
exit time by their standard deviation $\sigma(\Delta v)$ for each
given $\Delta v$. Then the PDF of the normalized exit times
$x=s/\sigma$ can be determined by
\begin{equation}
\rho(x) = \sigma f(x \sigma)~.
 \label{Eq:rho}
\end{equation}
Figure \ref{Fig:PDF} shows the empirical PDFs $\sigma f(x \sigma)$
of the normalized exit times $x = s / \sigma$ for different
thresholds $\Delta{v}$. The PDFs at different thresholds cannot be
superposed by the simple normalization procedure. As sketched in
Fig.~\ref{Fig:PDF}(a), the PDFs are strongly asymmetric and not
log-normal, which differs remarkably from the situation of energy
dissipation in three-dimensional fully developed turbulence showing
roughly log-normal distribution
\cite{Xu-Zhou-Liu-Gong-Wang-Yu-2006-PRE}. More interestingly, the
probability density functions show plateaus on the left tails, which
ensures the existence of any negative moments.
Figure~\ref{Fig:PDF}(b) shows that the right tail relaxes
exponentially for small thresholds or faster for large thresholds.
This relaxation behavior is different from those exit times
extracted from financial return series exhibiting a power-law tail
\cite{Simonsen-Jensen-Johansen-2002-EPJB,Jensen-Johansen-Petroni-Simonsen-2004-PA,Jensen-Johansen-Simonsen-2003-IJMPC,Jensen-Johansen-Simonsen-2003-PA,Zhou-Yuan-2005-PA,Simonsen-Ahlgren-Jensen-Donangelo-Sneppen-2007-EPJB},

\begin{figure}[htb]
\centering
\includegraphics[width=6.5cm]{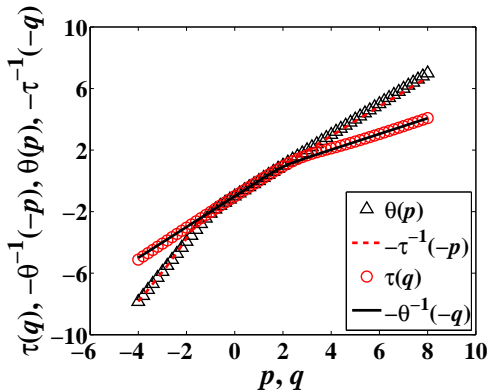}
\caption{Testing the inversion formula in financial volatility.}
\label{Fig:Matching}
\end{figure}

The power-law exponents $\tau(q)$ for direct partition functions are
plotted as open circles in Fig.~\ref{Fig:Matching}, while the
exponents $\theta(q)$ are illustrated as triangles. Both $\tau(q)$
and $\theta(p)$ are nonlinear, indicating that the time series of
volatility and exit time possess multifractal properties. The
function $-\theta^{-1}(-q)$ is determined numerically from the
$\theta(p)$ curve, which is plotted in Fig.~\ref{Fig:Matching} as a
dashed line. We can find that the values of $-\theta^{-1}(-q)$ are
in excellent agreement with the values of $\tau(p)$, which provides
strong evidence supporting the inversion formula in
Eq.~(\ref{Eq:IF1}). Similarly, the $-\tau^{-1}(-p)$ curve
numerically obtained from the $\tau(p)$ function is depicted as a
solid line, which coincides remarkably with the $\theta(q)$ curve.
In other words, the inversion formula Eq.~(\ref{Eq:IF2}) also holds
as expected. We note that the differences between the comparing
curves are well within the error bars.

In summary, we have attempted to test the inversion formula for
conservative multifractal measures using high-frequency volatility
data of the S\&P 500 index. We have performed multifractal analysis
on both volatility and exit time series based on the partition
function method. Our investigation confirms that both direct and
inverse partition functions exhibit nice multifractal properties.
The two scaling ranges are consistent with each other. Furthermore,
we found that the function $-\tau^{-1}(-p)$ extracted numerically
from $\tau(q)$ overlaps with the $\theta(p)$ curve and the function
$-\theta^{-1}(-q)$ determined from the $\theta(p)$ curve collapses
on the $\tau(q)$ curve, which verifies the inversion formula. We
also investigated for the first time the empirical distributions of
exit time of financial volatility at different thresholds. The PDFs
of exit time are nontrivial, which are neither log-normal nor power
laws observed in other systems.

\begin{acknowledgments}
We are grateful to Gao-Feng Gu for discussion. This work was partly
supported by the National Natural Science Foundation of China (Grant
No. 70501011), the Fok Ying Tong Education Foundation (Grant No.
101086), the Shanghai Rising-Star Program (Grant No. 06QA14015), and
the Program for New Century Excellent Talents in University (Grant
No. NCET-07-0288).
\end{acknowledgments}

%\bibliography{D:/Papers/Auxiliary/Bibliography} %Jiang
\bibliography{E:/Papers/Auxiliary/Bibliography} %Zhou

\begin{thebibliography}{31}
\expandafter\ifx\csname natexlab\endcsname\relax\def\natexlab#1{#1}\fi
\expandafter\ifx\csname bibnamefont\endcsname\relax
  \def\bibnamefont#1{#1}\fi
\expandafter\ifx\csname bibfnamefont\endcsname\relax
  \def\bibfnamefont#1{#1}\fi
\expandafter\ifx\csname citenamefont\endcsname\relax
  \def\citenamefont#1{#1}\fi
\expandafter\ifx\csname url\endcsname\relax
  \def\url#1{\texttt{#1}}\fi
\expandafter\ifx\csname urlprefix\endcsname\relax\def\urlprefix{URL }\fi
\providecommand{\bibinfo}[2]{#2}
\providecommand{\eprint}[2][]{\url{#2}}

\bibitem[{\citenamefont{Jensen}(1999)}]{Jensen-1999-PRL}
\bibinfo{author}{\bibfnamefont{M.~H.} \bibnamefont{Jensen}},
  \bibinfo{journal}{Phys. Rev. Lett.} \textbf{\bibinfo{volume}{83}},
  \bibinfo{pages}{76} (\bibinfo{year}{1999}).

\bibitem[{\citenamefont{Biferale et~al.}(1999)\citenamefont{Biferale, Cencini,
  Vergni, and Vulpiani}}]{Biferale-Cencini-Vergni-Vulpiani-1999-PRE}
\bibinfo{author}{\bibfnamefont{L.}~\bibnamefont{Biferale}},
  \bibinfo{author}{\bibfnamefont{M.}~\bibnamefont{Cencini}},
  \bibinfo{author}{\bibfnamefont{D.}~\bibnamefont{Vergni}}, \bibnamefont{and}
  \bibinfo{author}{\bibfnamefont{A.}~\bibnamefont{Vulpiani}},
  \bibinfo{journal}{Phys. Rev. E} \textbf{\bibinfo{volume}{60}},
  \bibinfo{pages}{R6295} (\bibinfo{year}{1999}).

\bibitem[{\citenamefont{Simonsen et~al.}(2002)\citenamefont{Simonsen, Jensen,
  and Johansen}}]{Simonsen-Jensen-Johansen-2002-EPJB}
\bibinfo{author}{\bibfnamefont{I.}~\bibnamefont{Simonsen}},
  \bibinfo{author}{\bibfnamefont{M.~H.} \bibnamefont{Jensen}},
  \bibnamefont{and} \bibinfo{author}{\bibfnamefont{A.}~\bibnamefont{Johansen}},
  \bibinfo{journal}{Eur. Phys. J. B} \textbf{\bibinfo{volume}{27}},
  \bibinfo{pages}{583} (\bibinfo{year}{2002}).

\bibitem[{\citenamefont{McCauley}(1990)}]{McCauley-1990-PR}
\bibinfo{author}{\bibfnamefont{J.~L.} \bibnamefont{McCauley}},
  \bibinfo{journal}{Phys. Rep.} \textbf{\bibinfo{volume}{189}},
  \bibinfo{pages}{225} (\bibinfo{year}{1990}).

\bibitem[{\citenamefont{Frisch}(1996)}]{Frisch-1996}
\bibinfo{author}{\bibfnamefont{U.}~\bibnamefont{Frisch}},
  \emph{\bibinfo{title}{Turbulence: The Legacy of A.N. Kolmogorov}}
  (\bibinfo{publisher}{Cambridge University Press},
  \bibinfo{address}{Cambridge}, \bibinfo{year}{1996}).

\bibitem[{\citenamefont{Anselmet et~al.}(1984)\citenamefont{Anselmet, Gagne,
  Hopfinger, and Antonia}}]{Anselmet-Gagne-Hopfinger-Antonia-1984-JFM}
\bibinfo{author}{\bibfnamefont{F.}~\bibnamefont{Anselmet}},
  \bibinfo{author}{\bibfnamefont{Y.}~\bibnamefont{Gagne}},
  \bibinfo{author}{\bibfnamefont{E.~J.} \bibnamefont{Hopfinger}},
  \bibnamefont{and} \bibinfo{author}{\bibfnamefont{R.~A.}
  \bibnamefont{Antonia}}, \bibinfo{journal}{J. Fluid Mech.}
  \textbf{\bibinfo{volume}{140}}, \bibinfo{pages}{63} (\bibinfo{year}{1984}).

\bibitem[{\citenamefont{Vandewalle and
  Ausloos}(1998)}]{Vandewalle-Ausloos-1998-EPJB}
\bibinfo{author}{\bibfnamefont{N.}~\bibnamefont{Vandewalle}} \bibnamefont{and}
  \bibinfo{author}{\bibfnamefont{M.}~\bibnamefont{Ausloos}},
  \bibinfo{journal}{Eur. Phys. J. B} \textbf{\bibinfo{volume}{4}},
  \bibinfo{pages}{257} (\bibinfo{year}{1998}).

\bibitem[{\citenamefont{Ivanova and Ausloos}(1999)}]{Ivanova-Ausloos-1999-EPJB}
\bibinfo{author}{\bibfnamefont{K.}~\bibnamefont{Ivanova}} \bibnamefont{and}
  \bibinfo{author}{\bibfnamefont{M.}~\bibnamefont{Ausloos}},
  \bibinfo{journal}{Eur. Phys. J. B} \textbf{\bibinfo{volume}{8}},
  \bibinfo{pages}{665} (\bibinfo{year}{1999}).

\bibitem[{\citenamefont{Calvet and Fisher}(2002)}]{Calvet-Fisher-2002-RES}
\bibinfo{author}{\bibfnamefont{L.}~\bibnamefont{Calvet}} \bibnamefont{and}
  \bibinfo{author}{\bibfnamefont{A.}~\bibnamefont{Fisher}},
  \bibinfo{journal}{Rev. Econ. Stat.} \textbf{\bibinfo{volume}{84}},
  \bibinfo{pages}{381} (\bibinfo{year}{2002}).

\bibitem[{\citenamefont{Schmitt}(2005)}]{Schmitt-2005-PLA}
\bibinfo{author}{\bibfnamefont{F.}~\bibnamefont{Schmitt}},
  \bibinfo{journal}{Phys. Lett. A} \textbf{\bibinfo{volume}{342}},
  \bibinfo{pages}{448} (\bibinfo{year}{2005}).

\bibitem[{\citenamefont{Hastings}(2002)}]{Hastings-2002-PRL}
\bibinfo{author}{\bibfnamefont{M.~B.} \bibnamefont{Hastings}},
  \bibinfo{journal}{Phys. Rev. Lett.} \textbf{\bibinfo{volume}{88}},
  \bibinfo{pages}{055506} (\bibinfo{year}{2002}).

\bibitem[{\citenamefont{Biferale et~al.}(2001)\citenamefont{Biferale, Cencini,
  Lanotte, Vergni, and
  Vulpiani}}]{Biferale-Cencini-Lanotte-Vergni-Vulpiani-2001-PRL}
\bibinfo{author}{\bibfnamefont{L.}~\bibnamefont{Biferale}},
  \bibinfo{author}{\bibfnamefont{M.}~\bibnamefont{Cencini}},
  \bibinfo{author}{\bibfnamefont{A.~S.} \bibnamefont{Lanotte}},
  \bibinfo{author}{\bibfnamefont{D.}~\bibnamefont{Vergni}}, \bibnamefont{and}
  \bibinfo{author}{\bibfnamefont{A.}~\bibnamefont{Vulpiani}},
  \bibinfo{journal}{Phys. Rev. Lett.} \textbf{\bibinfo{volume}{87}},
  \bibinfo{pages}{124501} (\bibinfo{year}{2001}).

\bibitem[{\citenamefont{Biferale et~al.}(2003)\citenamefont{Biferale, Cencini,
  Lanotte, and Vergni}}]{Biferale-Cencini-Lanotte-Vergni-2003-PF}
\bibinfo{author}{\bibfnamefont{L.}~\bibnamefont{Biferale}},
  \bibinfo{author}{\bibfnamefont{M.}~\bibnamefont{Cencini}},
  \bibinfo{author}{\bibfnamefont{A.~S.} \bibnamefont{Lanotte}},
  \bibnamefont{and} \bibinfo{author}{\bibfnamefont{D.}~\bibnamefont{Vergni}},
  \bibinfo{journal}{Phys. Fluids} \textbf{\bibinfo{volume}{15}},
  \bibinfo{pages}{1012} (\bibinfo{year}{2003}).

\bibitem[{\citenamefont{Roux and Jensen}(2004)}]{Roux-Jensen-2004-PRE}
\bibinfo{author}{\bibfnamefont{S.}~\bibnamefont{Roux}} \bibnamefont{and}
  \bibinfo{author}{\bibfnamefont{M.~H.} \bibnamefont{Jensen}},
  \bibinfo{journal}{Phys. Rev. E} \textbf{\bibinfo{volume}{69}},
  \bibinfo{pages}{016309} (\bibinfo{year}{2004}).

\bibitem[{\citenamefont{Beaulac and
  Mydlarski}(2004)}]{Beaulac-Mydlarski-2004-PF}
\bibinfo{author}{\bibfnamefont{S.}~\bibnamefont{Beaulac}} \bibnamefont{and}
  \bibinfo{author}{\bibfnamefont{L.}~\bibnamefont{Mydlarski}},
  \bibinfo{journal}{Phys. Fluids} \textbf{\bibinfo{volume}{16}},
  \bibinfo{pages}{2126} (\bibinfo{year}{2004}).

\bibitem[{\citenamefont{Pearson and van~de
  Water}(2005)}]{Pearson-vandeWater-2005-PRE}
\bibinfo{author}{\bibfnamefont{B.~R.} \bibnamefont{Pearson}} \bibnamefont{and}
  \bibinfo{author}{\bibfnamefont{W.}~\bibnamefont{van~de Water}},
  \bibinfo{journal}{Phys. Rev. E} \textbf{\bibinfo{volume}{71}},
  \bibinfo{pages}{036303} (\bibinfo{year}{2005}).

\bibitem[{\citenamefont{Zhou et~al.}(2006)\citenamefont{Zhou, Sornette, and
  Yuan}}]{Zhou-Sornette-Yuan-2006-PD}
\bibinfo{author}{\bibfnamefont{W.-X.} \bibnamefont{Zhou}},
  \bibinfo{author}{\bibfnamefont{D.}~\bibnamefont{Sornette}}, \bibnamefont{and}
  \bibinfo{author}{\bibfnamefont{W.-K.} \bibnamefont{Yuan}},
  \bibinfo{journal}{Physica D} \textbf{\bibinfo{volume}{214}},
  \bibinfo{pages}{55} (\bibinfo{year}{2006}).

\bibitem[{\citenamefont{Xu et~al.}(2006)\citenamefont{Xu, Zhou, Liu, Gong,
  Wang, and Yu}}]{Xu-Zhou-Liu-Gong-Wang-Yu-2006-PRE}
\bibinfo{author}{\bibfnamefont{J.-L.} \bibnamefont{Xu}},
  \bibinfo{author}{\bibfnamefont{W.-X.} \bibnamefont{Zhou}},
  \bibinfo{author}{\bibfnamefont{H.-F.} \bibnamefont{Liu}},
  \bibinfo{author}{\bibfnamefont{X.}~\bibnamefont{Gong}},
  \bibinfo{author}{\bibfnamefont{F.-C.} \bibnamefont{Wang}}, \bibnamefont{and}
  \bibinfo{author}{\bibfnamefont{Z.-H.} \bibnamefont{Yu}},
  \bibinfo{journal}{Phys. Rev. E} \textbf{\bibinfo{volume}{73}},
  \bibinfo{pages}{056308} (\bibinfo{year}{2006}).

\bibitem[{\citenamefont{Mandelbrot and
  Riedi}(1997)}]{Mandelbrot-Riedi-1997-AAM}
\bibinfo{author}{\bibfnamefont{B.~B.} \bibnamefont{Mandelbrot}}
  \bibnamefont{and} \bibinfo{author}{\bibfnamefont{R.~H.} \bibnamefont{Riedi}},
  \bibinfo{journal}{Adv. Appl. Math.} \textbf{\bibinfo{volume}{18}},
  \bibinfo{pages}{50} (\bibinfo{year}{1997}).

\bibitem[{\citenamefont{Riedi and
  Mandelbrot}(1997)}]{Riedi-Mandelbrot-1997-AAM}
\bibinfo{author}{\bibfnamefont{R.~H.} \bibnamefont{Riedi}} \bibnamefont{and}
  \bibinfo{author}{\bibfnamefont{B.~B.} \bibnamefont{Mandelbrot}},
  \bibinfo{journal}{Adv. Appl. Math.} \textbf{\bibinfo{volume}{19}},
  \bibinfo{pages}{332} (\bibinfo{year}{1997}).

\bibitem[{\citenamefont{Mantegna and Stanley}(2000)}]{Mantegna-Stanley-2000}
\bibinfo{author}{\bibfnamefont{R.~N.} \bibnamefont{Mantegna}} \bibnamefont{and}
  \bibinfo{author}{\bibfnamefont{H.~E.} \bibnamefont{Stanley}},
  \emph{\bibinfo{title}{{An Introduction to Econophysics: Correlations and
  Complexity in Finance}}} (\bibinfo{publisher}{Cambridge University Press},
  \bibinfo{address}{Cambridge}, \bibinfo{year}{2000}).

\bibitem[{\citenamefont{Jiang and Zhou}(2007)}]{Jiang-Zhou-2007-PA}
\bibinfo{author}{\bibfnamefont{Z.-Q.} \bibnamefont{Jiang}} \bibnamefont{and}
  \bibinfo{author}{\bibfnamefont{W.-X.} \bibnamefont{Zhou}},
  \bibinfo{journal}{Physica A} \textbf{\bibinfo{volume}{381}},
  \bibinfo{pages}{343} (\bibinfo{year}{2007}).

\bibitem[{\citenamefont{Jiang and Zhou}(2008)}]{Jiang-Zhou-2008-XXX}
\bibinfo{author}{\bibfnamefont{Z.-Q.} \bibnamefont{Jiang}} \bibnamefont{and}
  \bibinfo{author}{\bibfnamefont{W.-X.} \bibnamefont{Zhou}}
  (\bibinfo{year}{2008}), \bibinfo{note}{arXiv:0801.1710}.

\bibitem[{\citenamefont{Cizeau et~al.}(1997)\citenamefont{Cizeau, Liu, Meyer,
  Peng, and Stanley}}]{Cizeau-Liu-Meyer-Peng-Stanley-1997-PA}
\bibinfo{author}{\bibfnamefont{P.}~\bibnamefont{Cizeau}},
  \bibinfo{author}{\bibfnamefont{Y.-H.} \bibnamefont{Liu}},
  \bibinfo{author}{\bibfnamefont{M.}~\bibnamefont{Meyer}},
  \bibinfo{author}{\bibfnamefont{C.-K.} \bibnamefont{Peng}}, \bibnamefont{and}
  \bibinfo{author}{\bibfnamefont{H.~E.} \bibnamefont{Stanley}},
  \bibinfo{journal}{Physica A} \textbf{\bibinfo{volume}{245}},
  \bibinfo{pages}{441} (\bibinfo{year}{1997}).

\bibitem[{\citenamefont{Liu et~al.}(1999)\citenamefont{Liu, Gopikrishnan,
  Cizeau, Meyer, Peng, and
  Stanley}}]{Liu-Gopikrishnan-Cizeau-Meyer-Peng-Stanley-1999-PRE}
\bibinfo{author}{\bibfnamefont{Y.-H.} \bibnamefont{Liu}},
  \bibinfo{author}{\bibfnamefont{P.}~\bibnamefont{Gopikrishnan}},
  \bibinfo{author}{\bibfnamefont{P.}~\bibnamefont{Cizeau}},
  \bibinfo{author}{\bibfnamefont{M.}~\bibnamefont{Meyer}},
  \bibinfo{author}{\bibfnamefont{C.-K.} \bibnamefont{Peng}}, \bibnamefont{and}
  \bibinfo{author}{\bibfnamefont{H.~E.} \bibnamefont{Stanley}},
  \bibinfo{journal}{Phys. Rev. E} \textbf{\bibinfo{volume}{60}},
  \bibinfo{pages}{1390} (\bibinfo{year}{1999}).

\bibitem[{\citenamefont{L'vov et~al.}(1998)\citenamefont{L'vov, Podivilov,
  Pomyalov, Procaccia, and
  Vandembroucq}}]{Lvov-Podivilov-Pomyalove-Procaccia-Vandembroucq-1998-PRE}
\bibinfo{author}{\bibfnamefont{V.~S.} \bibnamefont{L'vov}},
  \bibinfo{author}{\bibfnamefont{E.}~\bibnamefont{Podivilov}},
  \bibinfo{author}{\bibfnamefont{A.}~\bibnamefont{Pomyalov}},
  \bibinfo{author}{\bibfnamefont{I.}~\bibnamefont{Procaccia}},
  \bibnamefont{and}
  \bibinfo{author}{\bibfnamefont{D.}~\bibnamefont{Vandembroucq}},
  \bibinfo{journal}{Phys. Rev. E} \textbf{\bibinfo{volume}{58}},
  \bibinfo{pages}{1811} (\bibinfo{year}{1998}).

\bibitem[{\citenamefont{Jensen et~al.}(2004)\citenamefont{Jensen, Johansen,
  Petroni, and Simonsen}}]{Jensen-Johansen-Petroni-Simonsen-2004-PA}
\bibinfo{author}{\bibfnamefont{M.~H.} \bibnamefont{Jensen}},
  \bibinfo{author}{\bibfnamefont{A.}~\bibnamefont{Johansen}},
  \bibinfo{author}{\bibfnamefont{F.}~\bibnamefont{Petroni}}, \bibnamefont{and}
  \bibinfo{author}{\bibfnamefont{I.}~\bibnamefont{Simonsen}},
  \bibinfo{journal}{Physica A} \textbf{\bibinfo{volume}{340}},
  \bibinfo{pages}{678} (\bibinfo{year}{2004}).

\bibitem[{\citenamefont{Jensen et~al.}(2003{\natexlab{a}})\citenamefont{Jensen,
  Johansen, and Simonsen}}]{Jensen-Johansen-Simonsen-2003-IJMPC}
\bibinfo{author}{\bibfnamefont{M.~H.} \bibnamefont{Jensen}},
  \bibinfo{author}{\bibfnamefont{A.}~\bibnamefont{Johansen}}, \bibnamefont{and}
  \bibinfo{author}{\bibfnamefont{I.}~\bibnamefont{Simonsen}},
  \bibinfo{journal}{Int. J. Modern Phys. B} \textbf{\bibinfo{volume}{17}},
  \bibinfo{pages}{4003} (\bibinfo{year}{2003}{\natexlab{a}}).

\bibitem[{\citenamefont{Jensen et~al.}(2003{\natexlab{b}})\citenamefont{Jensen,
  Johansen, and Simonsen}}]{Jensen-Johansen-Simonsen-2003-PA}
\bibinfo{author}{\bibfnamefont{M.~H.} \bibnamefont{Jensen}},
  \bibinfo{author}{\bibfnamefont{A.}~\bibnamefont{Johansen}}, \bibnamefont{and}
  \bibinfo{author}{\bibfnamefont{I.}~\bibnamefont{Simonsen}},
  \bibinfo{journal}{Physica A} \textbf{\bibinfo{volume}{324}},
  \bibinfo{pages}{338} (\bibinfo{year}{2003}{\natexlab{b}}).

\bibitem[{\citenamefont{Zhou and Yuan}(2005)}]{Zhou-Yuan-2005-PA}
\bibinfo{author}{\bibfnamefont{W.-X.} \bibnamefont{Zhou}} \bibnamefont{and}
  \bibinfo{author}{\bibfnamefont{W.-K.} \bibnamefont{Yuan}},
  \bibinfo{journal}{Physica A} \textbf{\bibinfo{volume}{353}},
  \bibinfo{pages}{433} (\bibinfo{year}{2005}).

\bibitem[{\citenamefont{Simonsen et~al.}(2007)\citenamefont{Simonsen, Ahlgren,
  Jensen, Donangelo, and
  Sneppen}}]{Simonsen-Ahlgren-Jensen-Donangelo-Sneppen-2007-EPJB}
\bibinfo{author}{\bibfnamefont{I.}~\bibnamefont{Simonsen}},
  \bibinfo{author}{\bibfnamefont{P.~T.~H.} \bibnamefont{Ahlgren}},
  \bibinfo{author}{\bibfnamefont{M.~H.} \bibnamefont{Jensen}},
  \bibinfo{author}{\bibfnamefont{R.}~\bibnamefont{Donangelo}},
  \bibnamefont{and} \bibinfo{author}{\bibfnamefont{K.}~\bibnamefont{Sneppen}},
  \bibinfo{journal}{Eur. Phys. J. B} \textbf{\bibinfo{volume}{57}},
  \bibinfo{pages}{153} (\bibinfo{year}{2007}).

\end{thebibliography}

\end{document}